\documentclass[11pt]{article}
\pdfoutput=1
\usepackage{graphicx,amssymb}
\begin{document}
\begin{center}\LARGE
	Regions of an excessive flux of cosmic rays\\ 
	according to data of the FIAN and MSU arrays
\end{center}

\begin{center}\large
	E.N. Gudkova$^1$, M.Yu.\ Zotov$^2$, N.N. Kalmykov$^2$, G.V.~Kulikov$^2$,\\
	N.M. Nesterova$^1$, V.P. Pavlyuchenko$^1$
\end{center}

\begin{center}
	1. P.N.~Lebedev Physical Institute of the Russian Academy of Sciences\\
	2. D.V.~Skobeltsyn Institute of Nuclear Physics,\\
	M.V.~Lomonosov Moscow State University\\[2mm]
	(To be published in the Proceedings of the 33rd Russian Cosmic Ray Conference,\\
	Dubna, 11--15 August, 2014)
\end{center}

\centerline{\textbf{Abstract}}
{\narrower\noindent
Results of a blind search for localised regions of an excessive flux of cosmic
rays in the energy range from 50~TeV to 20~PeV with the data of the FIAN
KLARA-Chronotron experiment, the EAS MSU array and the Prototype of the EAS-1000
array are presented. A number of regions with a significant excess of the
registered flux over an expected isotropic background are found. Some of the
regions are present in at least two of the data sets considered.

}

\section{Introduction}

Results of the Super-Kamiokande, Tibet-AS$_\gamma$, Milagro, IceCube and some other
experiments revealed localised regions of excess of primary cosmic rays (PCRs)
in the TeV-PeV energy range and thus attracted considerable attention because
the large-scale distribution of arrival directions of such PCRs is almost
isotropic, see the recent review~[1]. In the present work, we continue the
series of anisotropy studies of PCRs at medium angular scales basing on the data
of the KLARA-Chronotron experiment carried out at the Tian Shan station of the
Lebedev Physical Institute of the Russian Academy of Sciences (FIAN)
($43.04^\circ$N, $76.97^\circ$E, $P=690$~g~cm$^{–2}$) and two experiments
carried out at Lomonosov Moscow State University (MSU), one with the EAS MSU
array, another one with the prototype of the EAS-1000 array (PRO-1000 in what
follows) ($55.70^\circ$N, $37.54^\circ$E)~[2--9]. To the contrary to the previous
works, the data of all three experiments are analysed with a unified method,
namely the shuffling technique, which has demonstrated its effectiveness in
similar anisotropy studies. This allows us to compare the results basing on the
common ground.

\section{Main Results}

The KLARA-Chronotron experiment was carried out by FIAN together with the
Central Institute for Physics Research (KFKI) of the Hungarian Academy of
Sciences at a separate installation located at the Tian Shan scientific
station~[3]. The experiment was specifically designed for continuous studies of
anisotropy of PCRs and worked simultaneously but independently of the main FIAN
EAS array. Twenty three million events with energies from 50~TeV to 0.5~PeV with
zenith angles in the $20^\circ$--$60^\circ$ range registered in 1978--1982 were
selected for the present analysis. Half a million events registered with the
EAS MSU array in 1984--1990~[10] and 1.3 million events of PRO-1000
(1997--1999)~[11] were selected of the MSU data sets. 95\% of these events
correspond to PCRs in the energy range approximately 0.2--20~PeV in the first
case, and from 50~TeV to 5~PeV in the latter one. Thus, the energy range of
events registered with the FIAN array is fully covered by the PRO-1000 data.
The MSU events selected for the analysis have zenith angles $<45^\circ$. The
accuracy of arrival directions in all three experiments is estimated to be of
the order of $3^\circ$.

The analysis of the data was performed over the whole fields of view of the
experiments, namely for declination $\delta>-20^\circ$ for the FIAN experiment
and $\delta>10^\circ$ for those of MSU. The fields of view were covered with a
grid with $0.2^\circ\times0.2^\circ$ cells and scanned with circular regions of
different radii from $2^\circ$ to $8^\circ$. For the further analysis, we selected
regions in which the expected number of events exceeded 10000, 400 and 200 for
the FIAN, PRO-1000 and EAS MSU data sets respectively. The (pre-trial) statistical
significance~$S$ of deviation of the real number of events registered in a
particular region from the expected background was calculated with the formula
by Li and Ma, traditionally used in anisotropy studies~[12].

Regions found in the FIAN and MSU data sets such that the registered number of
events inside them exceeds the expected background by $S>3$ (regions with an
excessive flux, REFs in what follows) are shown in Fig.~1. For convenience,
the same regions of the celestial sphere are shown in all three cases. Numbers
in the upper panel (the FIAN data) mark nine regions that have a close
counterpart in at least one of the MSU data sets. Let us briefly discuss
coincidences that are the most interesting from our point of view.

\begin{figure}[!ht]
	\centerline{\includegraphics[width=.9\textwidth]{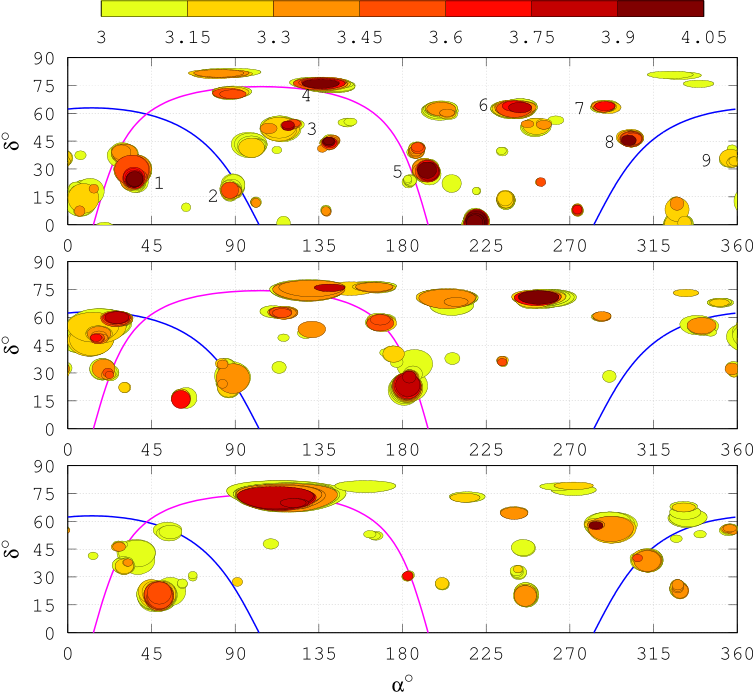}}
	\caption{From top to bottom: regions with an excessive flux of PCRs according
		to the FIAN, PRO-1000 and EAS MSU data sets respectively. Colours
		denote the (pretrial) statistical significance of deviation of the flux from the
		expected isotropic background. The equatorial coordinates are used. Curves at
		the left and right denote the Galactic plane, the $\cap$-like curve shows the
		Supergalactic plane.
	}
\end{figure}

Region 2 is located in the Galactic plane at
$\delta\approx14^\circ$--$28^\circ$. The maximum deviation from the expected
background flux is observed in a circle of radius $4.4^\circ$ centred at
($87.2^\circ$, $18.6^\circ$). 78061 events are registered inside the circle with
the expected 77061.2 events, which gives the pre-trial significance $S=3.6$. 
REF~2 intersects with an
extended region in the PRO-1000 data set ($\delta\approx18^\circ$--$38^\circ$) and
a small region in the EAS MSU data. This part of the celestial sphere is
interesting due to numerous potential sources of PCRs of TeV-PeV energies, among
them the Crab Nebula (SN1054), the supernova remnants IC443, PKS~0607+17, S147
and a number of energetic pulsars including Geminga~[13, 14], see Fig.~2.

\begin{figure}[!ht]
	\centerline{\includegraphics[width=.9\textwidth]{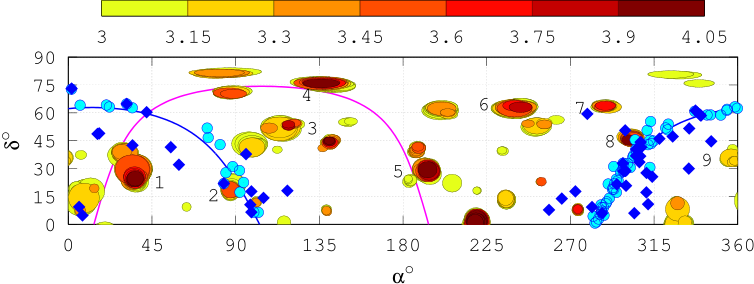}}
	\caption{Regions of an excessive flux of PCRs according to the FIAN data,
		Galactic supernova remnants (circles) according to~[13] and gamma-ray
		pulsars (diamonds) according to~[17].
	}
\end{figure}

Region 4 is located near the Supergalactic plane. In this case, the maximum
deviation from the isotropic flux almost reaches $S=4$. The EAS MSU and PRO-1000
data sets also demonstrate similar extended REFs, for which the excess of the
registered flux over the expected background exceeds 3.8 standard deviations.
Possible sources of PCRs with TeV--PeV energies are not known in this part of
the celestial sphere except for the close pulsar B0809+74 (0.43~kpc from the
Solar system), which is capable to accelerate protons up to $\sim4$~TeV, and pulsar
B0904+77, for which these values are not known at the moment~[14].

REF 8, which has the maximum deviation from the background flux in a circle of
radius $2.8^\circ$ centred at ($301.6^\circ$, $45.4^\circ$), also draws certain
interest. 26970 events are registered within this circle with the expected
number equal to 26327.2, which gives $S=3.9$. The region adjoins a REF in the EAS
MSU data set, which has a similar size. Both regions are located in the
direction to the Cygnus superbubble, which contains multiple supernova
remnants, energetic pulsars and OB-associations, see Fig. 2. Some of these
objects are located at distances $<1$~kpc from the Solar system. A cocoon filled
with PCRs accelerated up to energies $\sim0.3$~PeV was discovered by the Fermi-LAT
experiment approximately in this part of the sky~[15].

Region 7 is of particular interest since it has close REFs in both MSU data
sets, and an excess of the PCR flux over the expected background reaches
almost 4 standard deviations in the EAS MSU data~[7]. These regions are
remarkable in that they are located in the direction to the energetic
gamma-ray pulsar J1836+5925, which is similar to the Geminga pulsar and was
identified as such by the orbital Fermi-LAT experiment~[16, 17].

The FIAN data have also been studied by a method that takes into account a
dependence of absorption of extensive air showers in the atmosphere on the
zenith angle~[18].

\section{Conclusions}

There are a considerable number of localised regions with angular sizes up to
several dozen degrees with an excessive flux of PCRs in the energy range from
50~TeV to 20~PeV in the data sets obtained by the KLARA-Chronotron experiment
(FIAN) and with the EAS MSU and PRO-1000 arrays, and some of the regions found
in different data sets overlap or intersect.\footnote{The probability of
overlapping regions to appear by chance equals a product of the respective
probabilities for the data sets since all three sets were obtained
independently.}
As a whole, anisotropy of PCRs found with the FIAN data is closer to that for
PRO-1000, which is possibly due to the fact that both experiments covered close
energy intervals.

A direct comparison of the presented results with the results on anisotropy of
PCRs at similar angular scales obtained by other experiments is not
straightforward since the most pronounced REFs have been observed in the
1--10~TeV energy range but the picture considerably changes becoming more
mosaic-like as energy grows. One can notice an intersection of one of the REFs
in the FIAN data with an extended Region B found by the Milagro
experiment~[19]. All in all, the obtained picture of anisotropy is closer to
the results of Tibet-AS$_\gamma$ for energies 50~TeV and 300~TeV~[20].

It is difficult to give an unambiguous answer to the question about the origin
of the REFs but the regions that have close counterparts in two or all three
data sets are undoubtedly interesting, especially in case they are located in
parts of the celestial sphere with possible sources of PCRs of the considered
energies. Still, an existence of close supernova remnants or pulsars in this
case does not mean they are the reason of the REFs. In a $\sim1$~$\mu$G magnetic field,
a 1~PeV proton has a gyroradius of just 1~pc, which is much less than the
distance to any of the possible sources. Neutrons of these energies must also
be excluded from consideration because of the too short lifetime in free
state. The fraction of air showers initiated by gamma-quanta in this energy
range is likely to be negligibly small~[21, 22] and cannot lead to the
appearance of the discovered REFs. All this makes explaining the existence of
local inhomogeneities of the flux of TeV--PeV PCRs a rather complicated task.
At the moment, the most popular models are those based on the influence of
various configurations of Galactic magnetic fields,
see~[1] for a review.
There is no doubt that further studies of anisotropy of PCRs in the TeV--PeV
energy range are of considerable interest both for the cosmic ray physics and
for understanding the structure of the Galactic magnetic field.

The authors thank members of the staff of KFKI, MSU and the Tian Shan station
of FIAN who took part in constructing the arrays and performing the experiments.
The work was done with a financial support by the grant of the Government of
Russian Federation (contract 14.В25.31.0010) and grants of the Russian
Foundation for Basic Research 14-02-00372, 13-02-12175-ofi-m and 13-02-00214.

\section*{References}
\begin{enumerate}
	\item Di Sciascio G., Iuppa R.//Homage to the Discovery of Cosmic Rays, the
		Meson-Muon \& Solar Cosmic Rays, Ed. by Jorge A. Perez-Peraza, Nova Sci. Publ., N.Y. 2013.
	\item Benk\'o G. et al.//Izv.\ RAN, Ser.\ Fiz. 2004. V. 68. P. 1599.
	\item Benk\'o G. et al.//Nucl. Phys. B (Proc. Suppl.). 2008. V. 175. P. 541.
	\item Zotov M.Yu., Kulikov G.V.//Bull. Russ. Acad. Sci. Physics. 2004. V. 68. P. 1791.
	\item Zotov M.Yu., Kulikov G.V.//Bull. Russ. Acad. Sci. Physics. 2007. V. 71. P. 483.
	\item Zotov M.Yu., Kulikov G.V.//Bull. Russ. Acad. Sci. Physics. 2009. V. 73. P. 574.
	\item Zotov M.Yu., Kulikov G.V.//Astronomy Letters. 2010. V. 36. P. 645.
	\item Zotov M.Yu., Kulikov G.V.//Bull. Russ. Acad. Sci. Physics. 2011. V. 75. P. 342.
	\item Zotov M.Yu., Kulikov G.V.//Astron. Lett. 2012. V. 38. P. 731.
	\item Vernov S.N. et al.//Proc. 16th ICRC. Kyoto. 1979. V. 8. P. 129.
	\item Fomin Yu.A. et al.//Proc. 26th ICRC. Salt Lake City. 1999. V. 1. P. 286.
	\item Li. T.-P., Ma Y.-Q.//Astrophys. J. 1983. V. 272. P. 317.
	\item Green D.A.//Bull. Astron. Soc. India. 2009. V. 37. P. 45.
	\item Manchester R.N. et al.//Astron. J. 2005. V. 129. P. 1993.
	\item Ackermann, M. et al.//Science. 2011. V. 334. P. 1103.
	\item Abdo A.A. et al.//Astrophys. J. Suppl. Ser. 2009. V. 183. P. 46.
	\item Abdo A.A. et al.//Astrophys. J. Suppl. Ser. 2013. V. 208. P. 17.
	\item Gudkova E.N. et al.//Theses 33rd Russ. Cosmic Ray Conf. 2014. Dubna. P. 44.
	\item Abdo A. et al.//Phys. Rev. Lett. 2008. V. 101. P. 221101.
	\item Amenomori M. et al.//Science. 2006. V. 314. P. 439.
	\item Aglietta M. et al.//Astropart. Phys. 1996. V. 6. P. 71.
	\item Chantell M.C. et al.//Phys. Rev. Lett. 1997. V. 79. P. 1805.
\end{enumerate}
\end{document}